\documentstyle[PASJadd]{PASJ95}
\newcommand{\vol}{48}
\newcommand{\no}{6}
\newcommand{\lett}{}
\newcommand{\spage}{}
\newcommand{\rdate}{1996 July 8}
\newcommand{\adate}{1996 September 24}
\begin{document}
\title{Mass of the Galaxy Inferred from Outer Rotation Curve}
\author{Mareki {\sc Honma} and Yoshiaki {\sc Sofue}\\
{\it Institute of Astronomy, University of Tokyo, 2-21-1 Osawa, Mitaka, Tokyo, 181}\\
{\it E-mail(MH): honma@milano.mtk.nao.ac.jp}}
\abst{Using an outer rotation curve of the Galaxy, we explore the galactic constants and the mass of the Galaxy.
We show that $\Theta_{0}$ of 200 $\rm km\; s^{-1}\;$ is more favorable than the IAU standard value of 220 $\rm km\; s^{-1}$, and also show that if $\Theta_{0}$ is smaller than 207 $\rm km\; s^{-1}\;$ the rotation curve beyond 2$R_{0}$ is declining in Keplerian fashion.
In the case of $\Theta_{0}$= 200 $\rm km\; s^{-1}\;$ and $R_{0}$= 7.6 kpc, the total mass and the extent of the Galaxy inferred from the rotation curve are 2.0$\pm$0.3$\times$10$^{11}\MO$ and 15 kpc, respectively.
These results may significantly change the previous view of the Galaxy, that its outer region is dominated by a massive dark halo extending out to several tens of kpc.}
\kword{Galaxies: Milky Way --- Rotation --- Mass --- Dark Matter}
\maketitle
\thispagestyle{headings}

%section 1
\section{Introduction}
The mass of the Milky Way Galaxy has always been of greatest interest, but is still unknown.
Although a large number of studies have been performed to obtain the mass of the Galaxy based on the motions of globular clusters and satellite galaxies, the resultant mass varies by an order of magnitude from 10$^{11}\MO$ to 10$^{12}\MO$ (e.g., Zaritsky et al. 1989, and references therein).

As for extra-galaxies, recent high-sensitivity VLA observation have revealed that several galaxies do have a declining rotation curve in the outer region (e.g., Casertano, van Gorkom 1991), and that some of them are fitted well by a Keplerian (e.g., J\"ors\"ator, van Moorsel 1995; Olling 1996).
Therefore, if the Galaxy has a Keplerian rotation curve in its outer region, the outer rotation curve would enable us to measure the mass of the Galaxy directly.
Recently, Merrifield (1992) has developed a new method, and has derived the outer rotation curve of the Galaxy up to 2.5 $R_0$.
Based on this method, Honma and Sofue (1996) have investigated the outer rotation curve and its uncertainty.
In this paper, based on the outer rotation curve, we discuss the galactic constants and mass of the Galaxy.

%section 2
\section{Outer Rotation Curve and Galactic Constants}

In Honma \& Sofue (1996) we applied two independent methods proposed by Merrifield (1992) and by Petrovskaya and Teerikorpi (1986) to the {\sc Hi} survey data taken by Weaver and Williams (1974) and Kerr et al. (1986), and derived two corresponding rotation curves.
In addition to deriving rotation curves, we estimated the error of the derived rotation curves.
We found that the rotation curve derived by Merrifield's method is better than the other, having a smaller error and a wider coverage of the galacto-centric radius.
While the rotation curves previously obtained from the {\sc Hii} regions or stars do not extend beyond 2$R_0$, the rotation curve by Merrifield's method extends up to 2.5$R_0$, and its typical uncertainty is less than 30 $\rm km\; s^{-1}$, indicating that it is the most reliable one among the existing rotation curves at present.
In order to obtain a definite rotation curve, however, we need accurate galactic constants ($R_0$ and $\Theta_0$) because the Sun is orbiting around the galactic center.
For instance, we show in figure 1 the rotation curves for three cases of $\Theta_0$ ($\Theta_0= 180$ $\rm km\; s^{-1}$, 200 $\rm km\; s^{-1}\;$ and 220 $\rm km\; s^{-1}\;$).
The rotation curve for the inner Galaxy has been obtained from the tangential points in the {\sc Hi} position-velocity diagram (Fich et al. 1989).
%
% figure 1
%
This figure shows that the shape of the rotation curve is quite sensitive to $\Theta_0$, while different values of $R_0$ change only the scaling in the radial direction.
It is remarkable that the rotation velocity is decreasing beyond $2R_0$ for all cases, and is well fitted by a Keplerian if $\Theta_0$ is slightly smaller than 220 $\rm km\; s^{-1}\;$;
if we assume the Keplerian rotation velocity for four outermost points ($R\ge 2.11 R_0$), we obtain $\Theta_0$ of 207 $\rm km\; s^{-1}$,
and if we take the five outermost points ($R\ge 1.96 R_0$), we obtain $\Theta_0$ of 200 $\rm km\; s^{-1}$.
On the other hand, the rotation velocity within 2$R_0$, which is not decreasing, cannot be fitted by any means by a Keplerian.
If $\Theta_0$ is smaller than 200 $\rm km\; s^{-1}$, the rotation velocity decreases faster than a Keplerian.
We note that such a rotation curve appears if the mass distribution is flattened and sharply truncated.

At present, $\Theta_0$ of 220 $\rm km\; s^{-1}\;$ is recommended by IAU (see Kerr, Lynden-Bell 1986).
However, we show below that $\Theta_0$ is probably smaller.
If we take $\Theta_0= 220$ $\rm km\; s^{-1}$, the outer rotation curve is rising between $R$= 1.1$R_0$ and $2.0R_0$ by more than 50 $\rm km\; s^{-1}$.
As is widely known, spiral galaxies as luminous as ours have rotation curves which are almost flat within the optical disk (Rubin et al. 1985).
If we assume that the rotation curve of the Galaxy is also flat within $2R_0$, $\Theta_0$ should be reduced to 192 $\rm km\; s^{-1}$, which is slightly smaller than the value for a Keplerian rotation curve.

Another method to determine $\Theta_0$ is to measure the distance to the galactic center $R_0$.
Since $R_0$ is related to $\Theta_0$ by $A-B=\Theta_0/R_0$, where $A$ and $B$ are Oort's constants, and are determined fairly well (Kerr, Lynden-Bell 1986), a measurement of $R_0$ gives a good estimate of $\Theta_0$.
One of the popular methods to measure $R_0$ is to compare the kinematic distances of galactic objects calculated by a rotation curve with the real distances of the objects.
Our newly obtained rotation curve makes it possible to obtain an accurate kinematic distance, and, thus, an accurate value of $R_0$.
Using the distances of OH/IR stars in Moran (1993), which are directly determined by their geometry, we obtain $R_0$ = 7.7 kpc, while the same data give $R_0=8.9\pm0.8$ kpc if a flat rotation curve of 220 $\rm km\; s^{-1}\;$ is assumed (Moran 1993).
Using the distances of young stars obtained by CCD photometry (Turbide, Moffat 1993), we have obtained 7.9$\pm$1.0 kpc assuming no metallicity gradient in the outer Galaxy, and 7.2$\pm$1.0 kpc assuming a typical metallicity gradient (we excluded the data for $R>2R_0$ because of a lack of information concerning the metallicity gradient there).
These results of $R_0$ agree well with recent studies.
A direct measurement of the distance to Sgr B2 gives $R_0$ = 7.1$\pm$1.5 kpc (Reid et al. 1988).
The observation of masers in W49 gives $R_0$=8.1$\pm$1.1 kpc (Gwinn et al. 1992).
A large number of recent studies based on rather indirect methods also claim that $R_0$ is probably smaller than 8.0 kpc (see Reid 1993 for review).

If we naively average our results above, we obtain $R_0$=7.6 kpc, yielding $\Theta_0$=201 $\rm km\; s^{-1}\;$ using $A-B=26.4$ $\rm km\; s^{-1}\;$ kpc$^{-1}$ (Kerr, Lynden-Bell 1986).
This is slightly larger than what we found for a flat rotation curve, and is in good agreement with other results;
a comparison of the shape of the galactic rotation curve with other galaxies gives 200$\pm$10 $\rm km\; s^{-1}\;$ (Merrifield 1992), and a model analysis on the galactic disk also shows that $\Theta_0$ may be less than 200 $\rm km\; s^{-1}\;$ (Kuijken, Tremaine 1994).
Table 1 summarizes $\Theta_0$ so far derived by various methods.
The values of $\Theta_0$ obtained as mentioned above and claimed by recent studies are in remarkably good agreement with what we found for a Keplerian rotation curve.
%
% Table 1

% Section 3
\section{Mass of the Galaxy Inferred from Outer Rotation Curve}

Using the rotation curve and the galactic constants, we calculate the enclosed mass within radius $R$ using
\begin{equation}
M(R) = \alpha\frac{R\Theta^2}{G}.
\end{equation}
The coefficient $\alpha$ adjusts the flatness of the mass distribution, and is here taken to be unity, which corresponds to a spherical mass distribution.
In figure 2 we plot the enclosed mass within $R$ for $\Theta_0=180$, 200, and $220$ $\rm km\; s^{-1}$.
$R_0$ is taken so that $A-B=26.4$ $\rm km\; s^{-1}\;$ kpc$^{-1}$.
%
% figure 2
%
As we discussed above, $\Theta_0$ of 200 $\rm km\; s^{-1}\;$ is favorable, and in that case the rotation curve is declining in Keplerian fashion, indicating a truncation of the mass distribution.
This is clearly seen in figure 2; although the enclosed mass is increasing within 2$R_0$, it remains constant beyond that.
For that case, the inferred mass is 2.0$\pm 0.3 \times10^{11}\MO$ with a truncation of the mass distribution at $\sim 15 $kpc (using $R_0$ = 7.6 kpc).
We show the rotation curve in figure 3 with a Keplerian rotation curve for a point mass of 2.0$\times$10$^{11}\MO$.
%
% figure 3
%
The rotation curve beyond $2R_0$ is fitted fairly well by the Keplerian.

We compare the result with previous studies of the mass of the Galaxy.
The mass of the Galaxy has been estimated based on the motions of globular clusters and satellite galaxies (e.g., Lynden-Bell et al. 1983; Little, Tremaine 1987; Zaritsky et al. 1989).
While their resultant mass of the Galaxy varies by an order of magnitude, there have been studies which are consistent with our result;
Lynden-Bell et al. (1983) obtained a mass of 2.6$\pm0.8 \times10^{11}\MO$, and Little and Tremaine (1987) obtained a similar value of 2.4$^{+1.3}_{-0.7}\times10^{11}\MO$.
Although these studies are somehow model-dependent due to lack of a transverse velocity, and are based on the IAU standard galactic constants, they are fairly close to our result for $\Theta_0=200$ $\rm km\; s^{-1}$.
We note that the enclosed mass for $\Theta_0=220$ $\rm km\; s^{-1}\;$ in figure 2 reaches about $3.4\times10^{11}\MO$ at $2.5R$, which is close to the upper limit obtained by Lynden-Bell et al. (1983) and Little and Tremaine (1987).
Therefore, if the true value of $\Theta_0$ is 220 $\rm km\; s^{-1}\;$ and the mass obtained by Lynden-Bell et al. (1983) or Little and Tremaine (1987) is correct within error, the mass distribution must be truncated at a radius which is not far from 2.5$R_0$.
On the other hand, Zaritsky et al. (1989) obtained a mass larger than $10^{12}\MO$ based on a sample of galaxies which includes Leo I.
However, their result is highly sensitive to the most distant satellite, Leo I, which has a distance of $\sim$230 kpc and an extremely high radial velocity.
We note that their result is significantly reduced when they exclude Leo I (see their table 4 for detail).
Our result, which is comparable to their smaller mass, may suggest that Leo I is not bound by the Galaxy.

The enclosed mass for $\Theta_0=180$ $\rm km\; s^{-1}\;$ is decreasing beyond $2R_0$.
This result comes from the rotation velocity decreasing faster than Keplerian.
This is, of course, forbidden if the mass distribution is perfectly spherical.
However, if the mass distribution is not spherical, but flattened and truncated, then such a rotation curve is also allowed (e.g., Casertano 1983).
Therefore we will be able to constrain the shape of the mass distribution to some extent if the true value of $\Theta_0$ is found to be less than 200 $\rm km\; s^{-1}$.

% Section 4
\section{Discussion}

If the Galaxy has a declining rotation curve in Keplerian fashion, why don't other galaxies have such a rotation curve ?
As is widely known, the rotation curves of galaxies are generally flat (e.g, Bosma 1981; Rubin et al. 1985).
We point out that the Keplerian velocity decreases quite slowly in the outer region; in figure 3 the rotation velocity is still as high as 190 $\rm km\; s^{-1}$, even at $3R_0$.
That gentle slope of a Keplerian rotation curve might make it difficult to be discovered in other galaxies.
However, there are some galaxies known to have such a declining rotation curve (e.g., Sancisi, van Albada 1987; Casertano, van Gorkom 1991).
The latest VLA observations with high sensitivity have revealed other declining rotation curves (e.g., J\"ors\"ater, van Moorsel 1995; Olling 1996).
Casertano and van Gorkom (1991) suggested that the rotation curves of galaxies with a scale length smaller than 3.5 kpc are declining in the outer part.
Since the scale length of the Galaxy's disk is estimated to be between 3 kpc and 4 kpc (Merrifield 1992; Fux, Martinet 1995) using $R_0$= 8.5 kpc, the Galaxy probably belongs to such a category.
Moreover, J\"ors\"ater and van Moorsel (1995) found a Keplerian rotation curve in the outer region of a typical barred galaxy NGC 1365.
Therefore, it is not unlikely that the Galaxy, which is possibly a barred galaxy (e.g., Blitz, Spergel 1991), has a Keplerian rotation curve.

The optical disk of the Galaxy has a cutoff at $13 \sim 13.5$ kpc (Robin et al. 1992, $R_0$= 7.6 kpc is adopted).
In Figure 1 the rotation curves for all cases of $\Theta_0$ start declining to some extent at around $2R_0$ or 15 kpc with $R_0$= 7.6 kpc.
Therefore, there might be a relation between the cutoff of the optical disk and the decline of the rotation curve.
Van der Kruit and Searle (1982) studied the cutoff of the optical disk in edge-on galaxies.
Out of the seven galaxies which they studied, at least four indeed have a declining rotation curve in Keplerian fashion outside the cutoff of the optical disk.
The most prominent example is NGC 4244 (Olling 1996); others are NGC 891 (Sancisi, van Albada 1987; Rupen 1991) NGC 4013 (Bottema et al. 1987), and NGC 5907 (Casertano 1983; Sancisi, van Albada 1987).
Although their rotation curves do not totally rule out the possibility of an extended dark halo (e.g., Casertano 1983; Bottema et al. 1987), one of the simplest interpretations concerning them is that they are Keplerian.
In any case, it is noteworthy that they have a declining rotation curve outside the optical disk.
Therefore, the declining rotation curve of the Galaxy in the case of $\Theta_0=200$ $\rm km\; s^{-1}$ does not necessarily conflict with previous studies.

While the Keplerian rotation curve does not require dark matter beyond 2$R_0$, it is still necessary within 2$R_0$.
While an exponential disk has a rotation curve declining beyond 2.2$h$, the Galaxy's rotation velocity within 2$R_0$ is almost constant in the case of $\Theta_0=200$ $\rm km\; s^{-1}$, indicating the existence of dark matter within the optical disk.
However, the total amount of dark matter inferred from the rotation curve is significantly smaller than what has been previously expected from a flat rotation curve out to several tens of kpc.\par
\vspace{1pc}\par
M.H. is financially supported by the Japan Society for the Promotion of Science.

%\clearpage

\section*{References}
%\small

\re
Blitz L., Spergel D. N. 1991, ApJ 379, 631

\re
Bosma A. 1981, AJ 86, 1825

\re
Bottema R., Shostak G. S., van der Kruit P. C., 1987, Nature 328, 401

\re
Casertano S. 1983, MNRAS 203, 735

\re
Casertano S., van Gorkom J. H. 1991, AJ 101, 1231

\re
Gwinn C. R., Moran J. M., Reid M. J. 1992, ApJ 393, 149

\re
Fich M., Blitz L., Stark A. A. 1989, ApJ 342, 272

\re
Fux R., Martinet L. 1994, A\&A 287, L21

\re
Honma M., Sofue Y. 1996, PASJ submitted

\re
J\"ors\"ater S., van Moorsel G. A. 1995, AJ 110, 2037

\re
Kerr F. J., Bowers P. F., Jackson P. D., Kerr M. 1986, A\&AS 66, 373

\re
Kerr F. J., Lynden-Bell D. 1986, MNRAS 221, 1023

\re
Kuijken K., Tremaine S. 1994, ApJ 421, 178

\re
Little B., Tremaine S. 1987, ApJ 320, 493

\re
Lynden-Bell D., Cannon R. D., Godwin P. J. 1983, MNRAS 204, 87p

\re
Merrifield M. R. 1992, AJ 103, 1552

\re
Moran J. M. 1993, In Sub Arcsecond Radio Astronomy, ed R. J. Davis, R. S. Booth. (Cambridge University Press, Cambridge), p 62

\re
Olling R. P. 1996, AJ 112, 457

\re
Petrovskaya I. V., Teerikorpi P. 1986, A\&A 163, 39

\re
Reid M. J., Schneps M. H., Moran J. M., Gwinn C. R., Genzel R., Downes D., R\"onn\"ang B. 1988, ApJ 330, 809

\re
Reid M. J. 1993, ARA\&A 31, 345

\re
Robin A. C., Cr\'ez\'e M., Mohan V. 1992, ApJ 400, L25

\re
Rubin V. C., Burstein D., Ford W. K. Jr, Thonnard N. 1985, ApJ 289, 81

\re
Rupen M. P. 1991, AJ 102, 48

\re
Sancisi R., van Albada T. S. 1987, in Dark Matter in the Universe, IAU Symposium No. 117, ed J. Kormendy, G. R. Knapp (Reidel, Dordrecht) p67

\re
Turbide L., Moffat F. J. 1993, AJ 105, 1831

\re
van der Kruit P. C., Searle L. 1982, A\&A 110, 61

\re
Weaver H., Williams D. R. W. 1975, A\&AS 17, 1

\re
Zaritsky D., Olszewski E. W., Schommer R. A., Peterson R. C., Aaronson M. 1989, ApJ 345, 759

%\clearpage
\vspace{3pc}

% figure captions
\centerline{\bf figure captions}
{\bf Fig. 1.} Rotation curves of the Galaxy derived based on the geometry of the {\sc Hi} disk.
The filled circles are for $\Theta_0$= 220 $\rm km\; s^{-1}$, the open circles for $\Theta_0$= 200 $\rm km\; s^{-1}$, and the triangles for $\Theta_0$= 180 $\rm km\; s^{-1}$.
The radius is normalized by the distance to the galactic center $R_0$.
The error bars are almost the same for all cases.

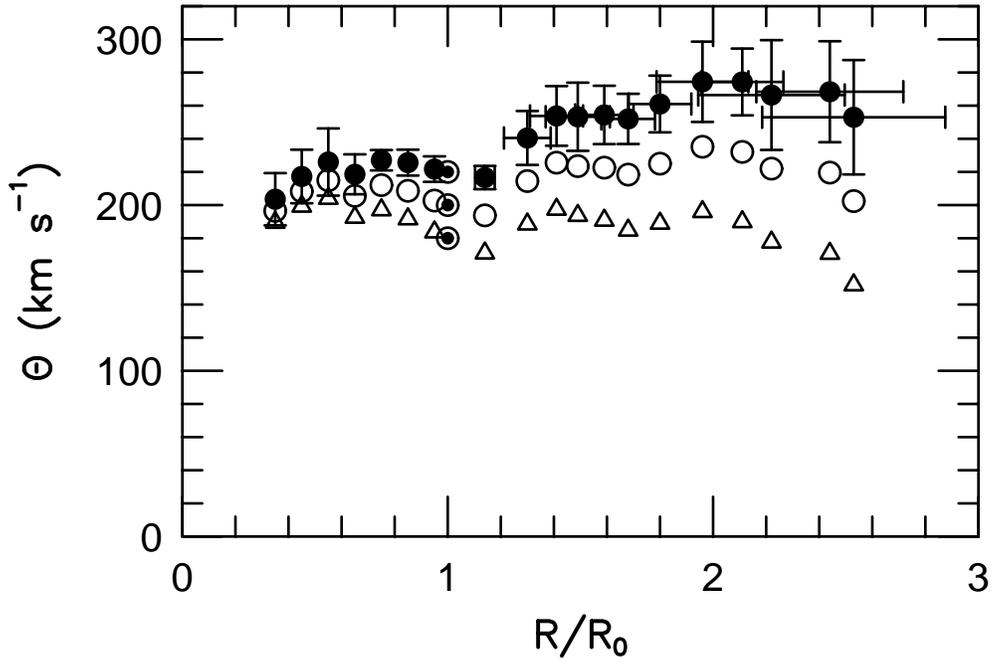
{\bf Fig. 2.} Enclosed masses within radius $R$ (equation 1).
$\Theta_0$ is taken to be 180, 200, and 220 $\rm km\; s^{-1}$, and $R_0$ is taken so that $A-B=26.4$ $\rm km\; s^{-1}\;$ kpc$^{-1}$.
The filled circles are for $\Theta_0$= 220 $\rm km\; s^{-1}$, the open circles for $\Theta_0$= 200 $\rm km\; s^{-1}$, and the triangles for $\Theta_0$=180 $\rm km\; s^{-1}$.

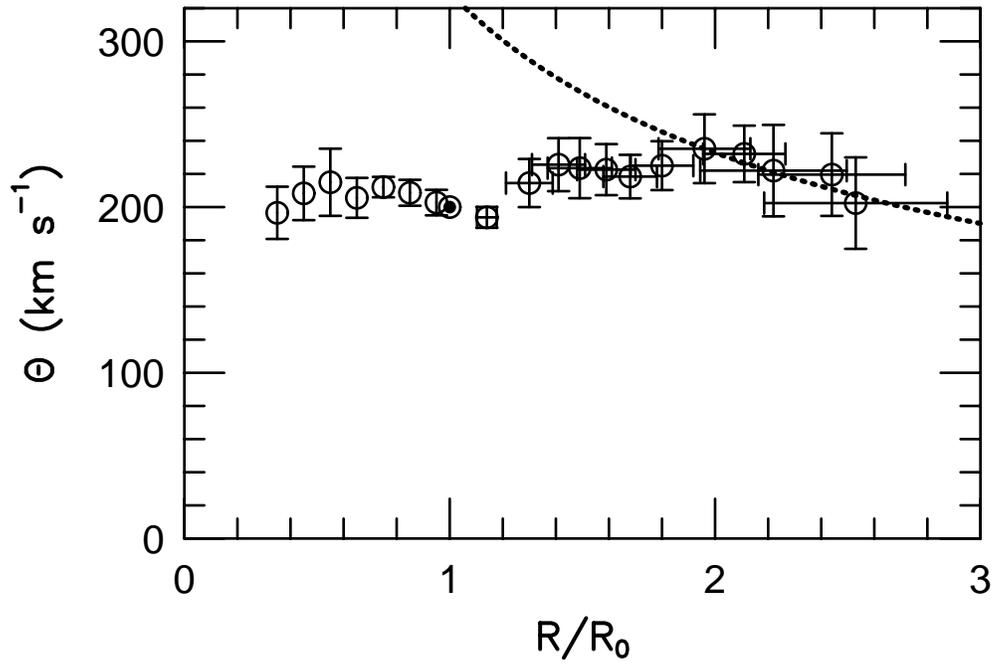
{\bf Fig. 3.} Rotation curve for $\Theta_0$= 200 $\rm km\; s^{-1}$, superposed on the Keplerian rotation velocity for a point mass of 2.0$\times$10$^{11}\MO$ (dotted line).
$R_0$ is taken to be 7.6 kpc.

%\begin{table*}[t]
%small
\vspace{3pc}

\begin{center}
Table~1.\hspace{4pt} $\Theta_0$ obtained by recent studies.\\
\vspace{6pt}
\begin{tabular}{ccc}
\hline\hline
$\Theta_0$ in $\rm km\; s^{-1}$ & References & Remarks\\
\hline
220 & Kerr and Lynden-Bell (1986) & IAU value \\
200$\pm$10 & Merrifield (1992) & Shape of rotation curve \\
180$\sim$200 & Kuijken and Tremaine (1994) & Model analysis \\
\hline
192 & This paper & For a flat rotation curve within 2$R_0$ \\
201 & This paper & From $R_0$=7.6 kpc and $A-B$=26.4 $\rm km\; s^{-1}$ kpc$^{-1}$\\
\hline
\end{tabular}
\end{center}
%\end{table*}

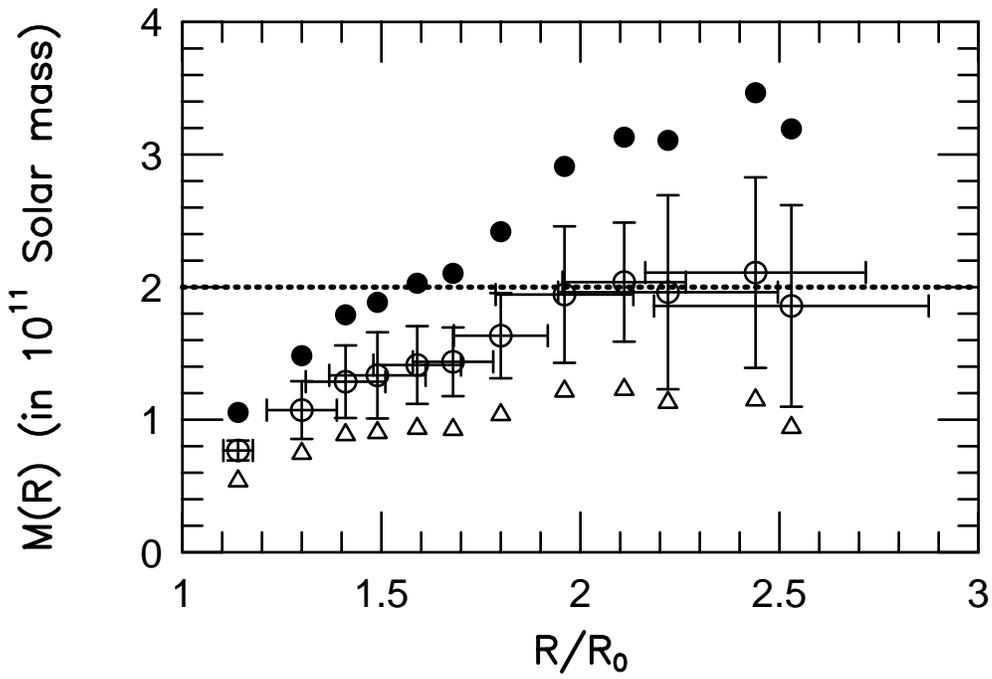
\end{document}